\documentclass[10pt,
aps,pra,
reprint,
notitlepage,
superscriptaddress,
frontmatterverbose, 
showpacs,
preprintnumbers,nofootinbib,
amsmath,amssymb,
]{revtex4-1}
\usepackage[combine]{ucs}
\usepackage[version=3]{mhchem} 
\usepackage{siunitx}
\usepackage{nicefrac}
\usepackage{graphicx}
\usepackage[british]{babel}
\usepackage{ifpdf}
\usepackage{natbib}
\usepackage{multirow}
\usepackage{appendix}
\usepackage{hyperref}

\hypersetup{pdfauthor={Salvador Rodríguez Gómez},pdftitle={Atomic charges for modeling metal--organic frameworks: Why and how}}
\begin{document}
\bibliographystyle{unsrtnat}
\preprint{}
\title{Atomic charges for modeling metal--organic frameworks: Why and how}
\author{Said \surname{Hamad}}
\email{said@upo.es}
\affiliation{Departament of Physical, Chemical and natural Systems, Universidad Pablo de Olavide, Ctra. Utrera km 1, 41013 Seville, Spain}
\author{Salvador \surname{R. G. Balestra}}
\affiliation{Departament of Physical, Chemical and natural Systems, Universidad Pablo de Olavide, Ctra. Utrera km 1, 41013 Seville, Spain}
\author{Rocío \surname{Bueno-Pérez}}
\affiliation{Departament of Physical, Chemical and natural Systems, Universidad Pablo de Olavide, Ctra. Utrera km 1, 41013 Seville, Spain}
\author{Sof\'{i}a \surname{Calero}}
\affiliation{Departament of Physical, Chemical and natural Systems, Universidad Pablo de Olavide, Ctra. Utrera km 1, 41013 Seville, Spain}
\author{A. Rabdel \surname{Ru\'{i}z-Salvador}}
\date{12th October 2016}
\begin{abstract}
Atomic partial charges are parameters of key importance in the simulation of Metal-Organic 
Frameworks (MOFs), since Coulombic interactions decrease with the distance more slowly than 
van der  Waals interactions.  But  despite its  relevance, there is  no method  to unambiguously 
assign  charges to  each atom,  since atomic  charges are  not  quantum  observables.  There are 
several methods that allow the calculation of  atomic charges, most of them starting from the 
wavefunction or  the electronic density  or the system,  as obtained with  quantum  mechanical 
calculations. In this work, we describe the most common methods employed to calculate atomic 
charges in MOFs. In order to show the influence that even small variations of structure have on 
atomic charges, we present the results that we obtained for DMOF-1. We also discuss the effect 
that small variations of atomic charges have on the predicted structural properties IRMOF-1. 
\end{abstract}
\maketitle
\section{Introduction}
Metal-Organic Frameworks have emerged as front-edge materials, due to their potential impact 
on  several types  of applications,  mainly those  based on  adsorption and  separation (such  as 
hydrogen  storage \cite{Suh2011},  methane  and  carbon  dioxide  capture \cite{eddaoudi2002systematic,sumida2011carbon},  or  hydrocarbon \cite{li2011metal} and 
enantiomeric separation \cite{li2011metal,Cychosz2010}). Unlike traditional nanoporous solids, i.e. zeolites, carbons, and
clays, MOFs do not only exhibit enormous surface areas (beyond 5000 \si{\metre\squared\per\gram}), but also a huge 
structural and compositional diversity, resulting from the large amount of research carried out, 
which has recently reached over 2000 scientific papers by year. Obviously, it is very expensive 
and  time  consuming  to  carry  out  experimental  studies  on  several  different  materials.  But 
computer modelling is a useful tool, which can help guiding the experimental search into new 
and potentially interesting materials. It is possible, for example, to use computer simulations to 
devise viable routes for materials selection, via large screenings \cite{Wilmer2011,Dubbeldam2012}.
Computer simulations can also provide a platform for understanding the material behavior at an atomic scale, which 
often leads to application-tailored materials design \cite{Thornton2012, R2010}.

Since the study of adsorption, separation and diffusion related phenomena involves the explicit
consideration of hundreds, or even thousands of atoms (particularly in structures with large unit
cells, such as MOFs) classical simulation methods are the first choice \cite{Dren2009,Reedijk2013}. It is worth noting
that recently, quantum mechanics-based calculations have emerged as valuable tools in this field \cite{Yu2013,sillar2009ab},
but in MOFs their computational cost still precludes its use for screenings of a larger
number of materials, for the calculation of adsorption isotherms, diffusion of complex
molecules, or the study of systems in which entropic effects are relevant, etc. In atomistic
classical simulations the energy of the system can be written as:
\begin{equation}
E=E_{\text{bonding}}+E_{\text{non-bonding}}
\end{equation}
where $E_{\text{bonding}}$ involves contributions directly related to bonded atoms, and are described by the
sum of bond, angles and dihedral terms, while $E_{\text{non-bonding}}$ includes the interactions between non-
bonded atoms and has the form:
\begin{equation}
E_{\text{non-bonding}}=E_{\text{van der Waals}}+E_{\text{coulombic}}
\end{equation}

The van der Waals interactions are usually described by the typical 12-6 Lennard-Jones potential:
\begin{equation}
E^{\text{LJ}}_{ij}=4\epsilon_{ij}\left[\left(\frac{\sigma_{ij}}{r_{ij}}\right)^{12}-\left(\frac{\sigma_{ij}}{r_{ij}}\right)^{6}\right]
\end{equation} where $\epsilon$ is the energy at the minimum and $\sigma$ is the distance at which the energy is zero. The Coulombic interactions are calculated as follows:
\begin{equation}
E^{\text{coulombic}}_{ij}=\frac{1}{4\pi\epsilon_0}\frac{q_iq_j}{r_{ij}}
\end{equation} where $r_{ij}$ is the distance between atoms $i$ and $j$, $q_i$ and $q_j$ are the corresponding atomic partial charges and $k_e=\nicefrac{1}{4\pi\epsilon_0}$ is the Coulomb's constant.

The parameters used for the calculation of bonded and van der Waals interactions are usually
taken from generic force fields, such as Dreiding \cite{SL1990}, UFF \cite{rappe1992uff}, OPLS \cite{jorgensen1988opls}, TraPPE \cite{MG1998,B1999,JM2004,MG1999} or
AMBER \cite{WD1992}. Lennard--Jones van der Waals interactions between different atoms are computed
using the Lorentz-Berthelot \cite{allen2017computer} or the Jorgensen mixing rules \cite{WL1986}. When specific molecules
force fields are used for modelling adsorbates, the atomic charges are usually taken from the
force field used. In a number of cases, however, using the generic or specific force fields the
experimental adsorption data are not reproduced, and hence transferable force field
parameterization is required, via fitting of parameters to reproduce experimental data \cite{Calero2004,MartnCalvo2012} or
via fitting to reproduce ab-initio surface energies \cite{Fang2014,Lin2014,Bureekaew2013}. The parameters that describe the van
der Waals interactions and the interactions between bonded atoms are usually employed directly
as taken from the generic force fields. But the atomic charges need to be calculated for each
material. Since the atomic charges arise from the electronic density of the solids, even small
chemical differences between related MOFs lead to differences in the charges, as was recently
shown for functionalized imidazolates \cite{Sevillano2012}.

For the computation of the intermolecular interactions (MOF-adsorbate and adsorbate--adsorbate
interactions), which control adsorption, diffusion and separation processes, it is important to
keep in mind that they are of non-bonded nature, and consequently their correct description
depends on achieving a balance between van der Waals and Coulombic contributions \cite{T2011}. This
implies that, if a generic force field is used, it is necessary to use charges that would be not very
different from those employed during the parameterisation of the force field. For example, the
parameters of the van der Waals interactions in the Dreiding and UFF force fields were fitted
employing Gasteiger \cite{J1980} and QEq charges \cite{AK1991}, respectively. This seems to be one of the main
reasons why calculated and experimental data do not agree, when generic force fields largely
fail to model intermolecular interactions. As illustration, \citet{Babarao2011} found that a good
agreement with experimental \ce{CO2} isotherms in ZIF-68 was obtained when ChelpG or Mulliken
charges were used in conjunction with the Dreiding force field.

The effect of the choice of the atomic charges on computing adsorption and diffusion properties
of MOFs has been a topic of increasing attention. A few years ago, \citet{Walton2008} showed that
the inclusion of the electrostatic interactions between adsorbate molecules and the framework
was crucial in reproducing the step-like adsorption of \ce{CO2} in IRMOF-1.
\citet{T2011} showed that even quadrupolar molecules, such as \ce{CO2}, can interact very distinctly with MOFs,
being the electrostatic interaction more or less relevant than the van der Waals interactions,
depending on the atomic charges employed. They found that the influence of the charges
on the adsorption properties is very material dependent, i.e. for some materials we observe the
same adsorption behavior, for a wide range of atomic charges, but for other materials, slight
changes in atomic charges produce large changes in the adsorption properties. They computed
\ce{CO2} adsorption isotherms up to 0.1 bar in IRMOF-1, ZIF-8, ZIF-90, and Zn(nicotinate)$_2$,
employing charges calculated by the REPEAT, DDEC, Hirshfeld and CBAC methods, and also
without considering charges. These methods exhibit significant differences in the values of the
charges that they predict, e.g. \ce{Zn} charges calculated with the mentioned methods in IRMOF-1
are 1.2787, 1.2149, 0.4229 and 1.5955, respectively. However, the adsorption isotherms are
very similar in Zn(nicotinate)$_2$, less similar in IRMOF-1 and ZIF-8 and very different in ZIF-90.

In a study with 20 different MOFs with different topologies, pore sizes, and chemical
characteristics, it was found that the guest--framework electrostatic interaction can account for
10--40\% of the \ce{CO2} uptake at very low pressure, and these values decrease at least by factor of
4 at high pressures, where guest--guest interactions dominate \cite{Zheng2009b}. \citet{Sevillano2012}
used three sets of framework charges, changing in a range of 30\% of their values, to examine its
effect on the adsorption of \ce{CO2} in ten ZIFs of different functionalities, and found that, while
adsorption heats are almost the same for ZIF-8 and small differences are observed for ZIF-96,
the effect of varying framework charges on ZIF-3, -7, -93 and -97 is large. The
hydrophobic character of ZIF-8 seems to be responsible for the negligible effect that the choice
of charges has on the values of \ce{CO2} adsorption heats, which is supported by the results of
\citet{Zhang2013}, who found that simulated methanol adsorption in ZIF-8 is not affected by the framework
charges.

When modelling water in MOFs, the choice of the charges is much more relevant. \citet{Castillo2008}
studied water adsorption in HKUST, and found that, in order to reproduce the experimental
adsorption isotherms in the low pressure range, the \emph{ab-initio} derived framework charges needed
to be scaled up by 25 \%. And \citet{Salles2011} studied the adsorption in the hydrophobic MIL-
47, finding that the previously used \emph{ab-initio} charges for modelling \ce{CO2} adsorption needed to be
scaled down by 30\%, in order to reproduce water adsorption behaviour.

The influence of the MOF framework charges on molecular diffusion has been a topic of less
research. The calculated self-diffusion coefficients for \ce{CO2} in ZIF-8 using charges obtained
with the CBAC, REPEAT, and DDEC, and ESP methods show significant differences \cite{Zheng2012}. The
latter set of charges provides results in good agreement with experimental values, but the other
three sets overestimate the diffusion coefficient between 1.5 and 20 times. \citet{Liu2010} used a
different set of charges (as well as different Lennard-Jones potentials), and the calculated self-
diffusion coefficient of \ce{CO2} in ZIF-8 was two times larger than in the previous cited work.
Since in a number of MOFs the proper choice of the framework charges is of key importance to
model correctly the adsorption and diffusion behaviour, it is natural that the simulation of
molecular separation would be also markedly influenced by the electrostatic interactions. For
instance, the simulated \ce{CO2}/\ce{CH4} selectivity in HKUST shows reverse behaviors when charges
are not considered at all than when there is a fully account of both host-guest and guest-guest
electrostatic interactions \cite{Yang2006}.
For quadrupolar molecules, such as \ce{CO2} and \ce{N2}, it has been observed that the atomic charges produce an electric field inside the nanopores that largely enhances the selectivity due to the difference in quadrupole moments \cite{Yang2007}.

In the following section we will present a brief description of the most used methods for
calculating atomic charges in MOFs, referring the reader to the relevant references for a more
in-depth description. Then, we will present the results of the calculations we have carried out, to
illustrate the influence of the structure on the charge calculation of DMOF-1. We will also show
how the different sets of framework charges predict different thermal behaviors of IRMOF-1.

\section{Methods for calculating atomic charges in MOFs}
There are several methods with which to calculate atomic charges. They are always developed
with the aim of providing the most realistic description of the system. But we have to take into
account the fact that atomic charges are not quantum observables. Electron density can be easily
calculated and studied, but, there are no operators to unambiguously determine the charges
associated to each atom. This makes the calculation of charges almost a matter of choice.
Nevertheless, there are several methods that can provide atomic charges which can be used to
model porous materials with reasonable accuracy. We will describe the most widely used
methods to calculate atomic charges, employing quantum mechanical calculations. Methods a)
and b) are based on the population analysis of the wavefunction, methods c), d), e), and f) are
based on the partition of the electron density, methods g), h), and i) are based on the fitting of
the electrostatic potential around the molecule, and methods i) and j) are semiempirical
approaches, the first based on electronegativity equalisation and the other on bond connection
sequences.

\subsection{Mulliken Charges}
Mulliken charges are obtained from the Mulliken Population Analysis \cite{Mulliken1955}. The first step in the
calculation of these charges is to obtain the wavefunction. Like in other methods, the partial
charge of atom $A$ ($q_A$) can be calculated as:
\begin{equation}
q_A=Z_A-\int_{V_A}{\rho_A(r)dr}
\end{equation} 
where $Z_A$ is the charge of the positively charge atom core, and $\rho_A(r)$ is the electron density
surrounding the core, associated to that atom. This seemingly simple equation becomes very
complex when we want to know which part of the total electron density (which can be easily
calculated with any quantum mechanical calculation) is associated to that particular atom. And
here is where each method makes a different choice. In the Mulliken method the charge is
calculated as:
\begin{equation}
q_A=Z_A-G_A
\end{equation}
where $G_A$ is the gross atom population for atom $A$, which is calculated as the sum of the
population of all orbitals belonging to atom $A$. The population matrix is constructed by
assigning half the electron density to each of the two atoms that share electrons in a bond,
regardless of the electronegativity of the atoms.

Mulliken charges have been widely used, mainly due to the simplicity and computational speed
with which they can be obtained. For these reason they have been widely used in MOFs \cite{Ramsahye2007,Szeto2007,Watanabe2009,Rydn2013,Gaponik2005,Messner2013,Bao2011,Khvostikova2010,Xiong2010,Yang2014}.
There are two main problems with the Mulliken charges. Firstly, they are very dependent on the
molecular geometry and the basis set, so that small changes in either the geometry or the basis
sets give rise to large differences in the calculated charges \cite{Martin2004}. And secondly, they do not
provide a good description of the degree of covalency in bonds.

\subsection{Natural Population Analysis charges}
In order to overcome the problems associated with the Mulliken method, \citet{Reed1985}
developed the Natural Population Analysis (NPA). NPA charges are calculated using a set of
orthonormal orbitals called natural atomic orbitals (NAOs), which are generated from the
atomic orbitals that form the basis set. NAOs are used to calculate another set of orthonormal
orbitals, called natural bond orbitals (NBOs), which are then used to perform the population
analysis that provides the NPA charges. NPA charges usually provide charges that are not very
dependent on the molecular conformation or the basis set, but they have not been developed to
be calculated on periodic systems, so that the cluster approach (see section g) must be used if
the charges of a periodic system need to be calculated. That is one of the main reasons why they
have not been used often in the study of MOFs. Nevertheless there are some studies in which
they have been used \cite{Sevillano2012, Xiong2010, Choomwattana2008, Noro2009}.

\subsection{Bader charges}
These charges are calculated using Bader's atoms-in-molecules (AIM) theory. In this theory it is
possible to partition the electron density and assign the density to each atom, by analysing the
gradient and the Laplacian of the electron density. The electron density must be obtained first,
using any quantum mechanical method (HF, post-HF, DFT, etc.). Once we have the electron
density, we look for critical points in the middle of each bond, which are the points along the
line between two atoms at which the electron density is minimal. From that point a surface is
created by moving along the direction given by the gradient vector (that points to the direction
of fastest electron density decrease). This gradient vector will creates a surface that encloses a
certain volume, which will be the volume associated to the atom enclosed. The integral of the
electron density within that volume will provide the negative charge of the atom, and the partial
charge is the atom can be calculated just by subtracting that negative charge to the positive
charge of the nucleus.

Despite being useful to provide atomic partial charges, this method has been more frequently
used to get information about the changes on the electron density that take place upon
adsorption \cite{Yang2013} or the differences in electron density when the metals sites of MOFs are
changed \cite{Canepa2013}. Direct use of Bader charges in MOFs is not found very often \cite{Yang2013, Yang2012,Jrgensen2012,Tabrizi2014,Zhou2011,Park2012}.

\subsection{Density Derived Electrostatic and Chemical (DDEC) charges}
This method (developed by \citet{Manz2012}), is based on the atoms-in-molecules method
described above, but there are two main differences: it is designed to incorporate spherical
averaging to minimise atomic multipole magnitudes (in order to get a better description of the
electrostatic potential) and it uses reference ion densities to enhance the transferability and
chemical meaning of the charges.

These charges are better suited to model porous materials than Bader charges, because the latter
do not give a correct description of the electrostatic potential (because they predict too large
atomic multipole moments \cite{Manz2010}). This method has been used to study the adsorption of water in
Cu-BTC \cite{Zang2013}, \ce{N2}/\ce{CO2} separation in a large number of MOFs \cite{Haldoupis2012}, \ce{CO2}/\ce{CH4}, \ce{CO2}/\ce{N2} and
\ce{H2}/\ce{CO2} separations in several MOFs \cite{Erucar2014} and separation in \ce{Zr}-Based MOFs \cite{Jasuja2012}.

\subsection{Hirshfeld charges}
In the Hirshfeld method \cite{Hirshfeld1977} the population of each atom is calculated by assuming that the
charge density at each point is shared among the surrounding atoms in direct proportion to their
free-atom densities at the corresponding distances from the nuclei. There have been several
improvements upon the original Hirshfeld scheme, such as the Iterative Hirshfeld \cite{Bultinck2007} method
(HI), Fractional Occupation Hirshfeld-I method (FOHI-D) \cite{Geldof2014}, and the Extended Hirshfeld
method (HE) \cite{Verstraelen2013}, which has been proved to provide good results for periodic materials \cite{Vanpoucke2012}.
Hirshfeld charges has been used for the development of MIL-53(\ce{Al}) force field \cite{Vanduyfhuys2012} and
modelling functionalizing effects in MIL-47 \cite{Biswas2013}, among other works.

\subsection{Charge Model 5 (CM5) charges}
This method was developed by \citet{Marenich2012} and it uses the charges obtained from a
Hirshfeld population analysis (of a wavefunction obtained with density functional calculations)
as a starting point. The charges are then varied, using a set of parameters derived by fitting to
reference values of the gas-phase dipole moments of 614 molecular structures. CM5 charges
have been successfully used to study hydrocarbon separation \cite{Verma2013} and \ce{N2}/\ce{CH4} separation in
MOF-74 with various types of metal atoms \cite{Lee2013}. These charges can also be used to study the
hydration of molecules in aqueous solutions, obtaining the best results when the charges are
scaled by the factor 1.27 \cite{Vilseck2014}.

One drawback of this method is that it is implemented as a script that uses the output from
the Gaussian09 code as the input for calculating the charges. This means that only non-periodic
systems can be studied, and the calculation of charges of periodic systems must be performed
making use of the cluster approach, which is explained in the following method.

\subsection{Electrostatic Potential (ESP) derived charges}
\begin{figure}[!h]
	\begin{center}
		\centering
	    \includegraphics[width=0.48\textwidth]{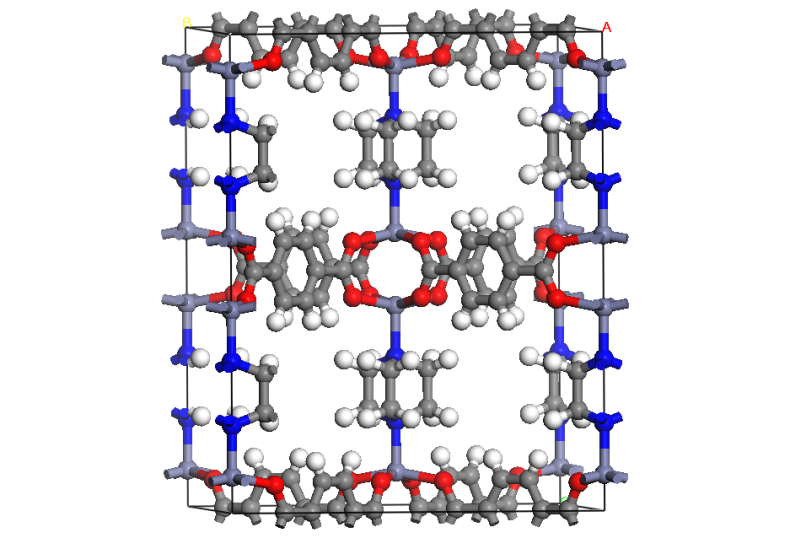}\\
	    \includegraphics[width=0.48\textwidth]{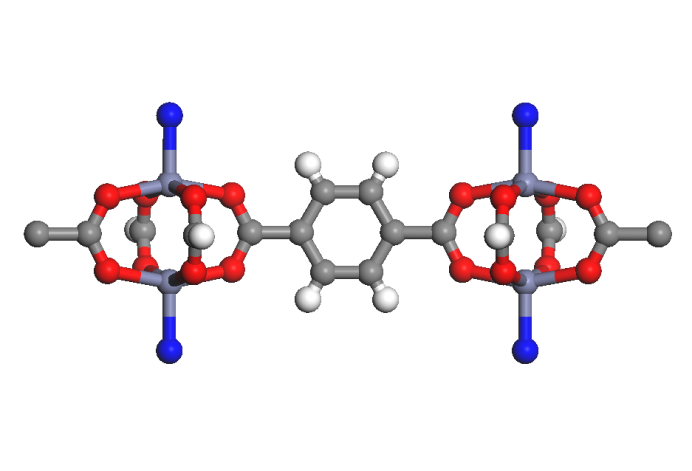}\\
	    \includegraphics[width=0.48\textwidth]{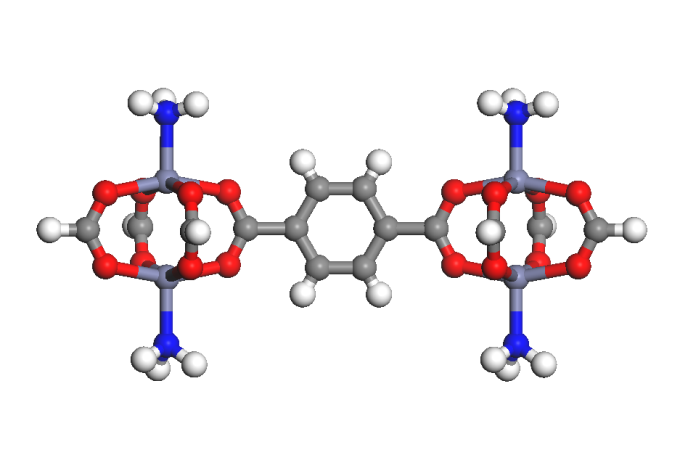}
	    \caption{Top) Ball and stick representation of the atoms of the unit cell of DMOF-1 (Zn, O, N, C, and H atoms are represented as light blue, red, dark blue, grey and white atoms respectively). Middle) Cluster created by cutting directly a piece of framework. This cluster cannot be used to model the environment of the BDC ligand and calculate its charges, since there are cleaved bonds that will have very different electronic structures than in the bulk structure. Bottom) Same cluster shown in b), although the cleaved bonds have been saturated with H atoms in order to achieve electronic structures in the terminal N and C atoms that are similar to those in the crystal structure.}
	    \label{fig:DMOF1_clusterb}
	\end{center}
\end{figure}
The first step is the calculation of the electrostatic potential around the molecule of interest,
using any quantum mechanical method. Once this potential is known for each point of space, a
set of initial atomic charges is assigned to each atom. With these initial charges, the potential on
a grid of points placed in a surface around the molecule is calculated, and an iterative method is
followed with which to fit the atomic partial charges that minimise the difference between the
quantum mechanical ESP and the one calculated with the atomic partial charges. There are
various methods to calculate ESP charges (differing in the choice of the points at which to
calculate the potential), such as CHELPG (CHarges from Electrostatic Potentials using a Grid-
based method \cite{Breneman1990}) and Merz-Kollman \cite{Singh1984}. The main drawback of these methods is that they
allow the calculation of charges for non-periodic systems. For crystals these methods cannot be
applied, since the electroscatic potential in periodic systems is not uniquely determined, because
there is a constant shift at each point of space. This problem has been circumvented by using the
so-called cluster approach (see Figure \ref{fig:DMOF1_clusterb}). This approach consists in using a cluster model of the
crystal, i.e. cutting a piece of crystal bulk, in the hope that the ESP derived charges for this
cluster model will be the same than for the bulk. This approximation works better for larger
clusters, so usually the charges are calculated for clusters of different sizes, until convergence is
achieved. There is another drawback for these methods, which is the fact that when the crystal is
cut to create the cluster model, there will be several bonds cleaved, leaving dangling bonds.

They are usually saturated with H atoms or with methyl groups. But these species are not part of
the original crystal, and they might have an influence on the fitted charges. Nevertheless, ESP
derived charges, have been the most widely used methods to obtain atomic partial charges, with
large success in modelling MOFs \cite{Ramsahye2007,Xiong2010,Vanduyfhuys2012,Sagara2004,Tafipolsky2007,Chen2010,Grosch2012,Zhang2012,Sun2011,Ma2012,Qiao2014}.
Only in the last few years they have been gradually replaced by other methods better suited for studying periodic systems.

\subsection{Repeating Electrostatic Potential Extracted Atomic (REPEAT) charges}
This method is similar to the ESP based methods described above. It was developed by \citet{Campana2009},
with the aim of solving the problems that ESP methods presented in the
study of periodic systems. The key point is the introduction of an error functional which acts on
the relative differences of the potential and not on its absolute values. For non-periodic systems
the REPEAT method provides charges that are very similar to those obtained with the CHELPG
method, and for periodic systems the charges it provides are chemically sound. Another
advantage of REPEAT charges is that is predicts similar charges when different codes (such as
CPMD, VASP or SIESTA) are employed. This method is becoming very popular to model
MOFs \cite{Jasuja2012,Ray2012,Vaidhyanathan2010,Morris2010,Morris2012,Sutrisno2012}.

\subsection{Density Derived Atomic Point (DDAP) charges}
This method was developed by \citet{Blochl1995}. It is based on the use of plane-waves to
calculate the density of a molecule. Atom-centered Gaussians are used to decouple the density
of the molecule (or each portion of the structure) from its periodic images, and the Ewald
summation is used to calculate their interaction energy. Finally, the charge density is modelled
with a set of atomic point charges. Although these charges can be used to study MOFs, its main
used has been in the study of ionic liquids \cite{Schmidt2010,Dommert2013,Zhang2012b}.

\subsection{Extended Charge Equilibration (EQEq) charges}
This method is based on the Charge Equilibration (QEq) method of \citet{AK1991}.
In the QEq method the charges are calculated using a set of experimental data, namely
atomic ionisation potentials, electron affinities, and atomic radii, with which an atomic chemical
potential is obtained (taking also into account shielded electrostatic interactions between all the
atomic charges). These charges are iteratively changed, until the equilibrium is found, when the
chemical potentials are equal in all atoms. The EQEq method \cite{CE2012} uses less fitting parameters,
while maintaining the accuracy. One important aspect in the charge equilibration methods is that
they do not require the calculation of wavefunction of electron densities; the only data needed
are the positions of the atoms and their atomic number. For this reason, these are the fastest
methods in terms of computation time, which makes them very useful for performing screenings
of a large number of materials \cite{CE2012,Wilmer2012b,Bernini2014}. Recent reparametrisations of the Qeq method have
been carried out by \citet{Haldoupis2012} and by \citet{Kadantsev2013}.

\subsection{Connectivity-based atom contribution (CBAC) charges}
In this method (developed by \citet{Xu2010}) there is no need to perform quantum
calculations, as happened in the EQEq method. The basis of this method was the assumption
that atoms with same bonding connectivity have identical charges in different MOFs. They first
obtained the charges of a set of 30 MOFs, using the ChelpG method (with the cluster approach)
from the electron density calculated with unrestricted B3LYP calculations. The basis set
employed is LANL2DZ for the metal atoms and 6-31+G* for the rest. The average charges for
similar atoms were calculated and tabulated. It is therefore possible to obtain the charges of any
MOFs, as long as it has the same types of atoms that were studied in the set of 30 MOFs (plus
16 COFs with which the database was subsequently expanded \cite{Zheng2010}). There is one small
drawback associated with the wide range of MOFs that can be studied with this method, which
is that in some cases the structures are not charge neutral. Nevertheless, it is very easy to
calculate charges with this method, and they usually provide good results, so they are frequently
used to model adsorption in MOFs \cite{Robinson2012,Zhuang2011}.

\section{Learning from two examples}
Here we show two examples chosen to illustrate two different aspects related to the charges, i.e.
(a) influence of the framework geometries on the calculated charges and (b) influence of the
chosen charges on structural properties, namely the negative thermal expansion of IRMOF-1.

\subsection{Influence of the framework geometries on the calculated charges}
\begin{figure*}[!htp]
    \begin{center}
    \centering
	    \includegraphics[width=0.6\textwidth]{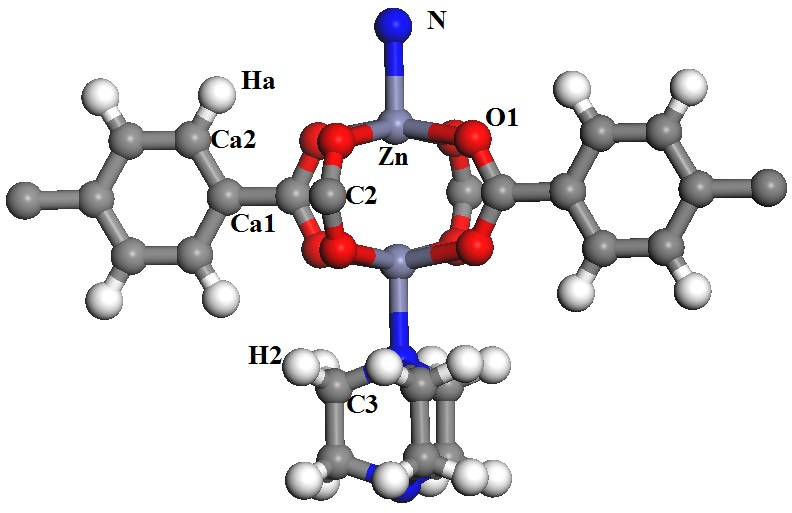}
	    \caption{Atom labels of DMOF-1 (see Tables 1, 2, and 3 for charges associated to the C3, Zn, and H2 atoms respectively).}
	    \label{fig:DMOF1_clusterc}
	\end{center}
\end{figure*}

We have calculated atomic charges of DMOF-1, which exhibits breathing-like flexibility \cite{Dybtsev2004}.
The dabco pillars are disordered along the fourfold crystallographic axis. Such disorder
precludes the direct use of the structure for calculating the charges, due to the atoms overlap.
Thus, the reported crystal structure in the I4/mcm (\# 140) space group needs to be fixed for its
description without symmetry (using a P1 space group). The obtained P1 structure has a number
of constrained bonds that can be relaxed using a generic force field (we employed the UFF force
field in our case). This structure is labelled as DMOF-1-ini. We have labelled as DMOF-1-opti1
the structure after an optimisation has been carried out at the DFT-D level, with the VASP code
\cite{Kresse1993,Kresse1994}. The dabco unit has a complex structure and their atoms in the DMOF-1-ini structure
are slightly disordered, the optimisation leads to a configuration with a relatively high energy.
For that reason, we carried out an additional optimisation, in which we first adjusted the
symmetry of the system, and then we reoptimise it with VASP. We called this third structure
DMOF-1-opti2. All the VASP calculations are carried out employing the PAW potentials \cite{PhysRevB.59.1758},
with the PBE exchange-correlation functional \cite{PhysRevLett.77.3865}, and a cut-off energy of 500 eV. Due to the
large sizes of the unit cells ($a=15.0630$ \AA and $c=19.2470$ \AA) only the gamma point was used.
The framework of DMOF-1 is shown in Figure \ref{fig:DMOF1_cluster}a, while the atom labels used for reporting the charges are shown in Figure \ref{fig:DMOF1_clusterc}.

The calculations with the VASP code permit the calculation of the REPAT and Bader charges,
for which we use the codes provided by \citet{Campana2009} and \citet{Tang2009} respectively.
We also calculated the Mulliken and DDAP charges for the same structures, using the cp2k code \cite{Laino2006} and
the PBE exchange-correlation functional. Finally, we also calculated the EQeq charges, with the code provided by \citet{Wilmer2012b}.
We report, in Table \ref{tbl:tb1_charges}, the range of variation of the charges of the C3 atom, for the three studied structures. We can see that, overall,
the values of the calculated charges vary in a very wide range. For a given structure, each
method provides different charges, ranging for instance the charge of the C3 atom, in the
DMOF-1-ini structure, from $-0.299$ to $+0.128$ when calculated with the REPEAT method, while
when the DDAP method is used the range of variation goes from $+0.224$ to $0.240$. This is not
surprising, since we have mentioned the intrinsic subjectivity associated to the process of
assigning the electronic density to each atom. The smaller range of variation is observed for the
Mulliken method, while the Bader charges are the ones that show a larger range of variation. A
similar behavior is observed for the \ce{Zn} and \ce{H2} atoms, as can be seen in Tables \ref{tbl:tb2_charges} and \ref{tbl:tb3_charges}. The
tables with the charges of the rest of the atoms are presented in the Supporting Information. The
large range of variation of the charges in structures DMOF-1-ini and DMOF-opti1 indicates that
the obtained charges will not be able to be used in force field-based simulations, since atoms
that should have the same chemical behavior are predicted to have very different charges. It is
worth noticing that three of the methods (Bader, DDAP and EQeq) predicted a negative charge
for the \ce{H2} hydrogen atom (see Table \ref{tbl:tb3_charges}).
\begin{table}[!h]
   \caption{Range of variation of the atomic partial charges for atom C3, calculated for the structures DMOF-1-ini, DMOF-1-opti1 and DMOF-1-opti2, using 5 different methods, namely REPEAT, Bader, Mulliken, DDAP, and EQeq.}
   \centering
  \label{tbl:tb1_charges}
  \begin{tabular}{cccc}
    \hline
    Method &  DMOF-1-ini & DMOF-1-opti1 & DMOF-1-opti2\\
    \hline  
    REPEAT & -0.299; 0.128 & -0.573; 0.308 & -0.363;	0.137 \\
    Bader & -0.03; 0.443 & 0.176; 0.543 & 0.187;	0.598 \\
    Mulliken & -0.052; -0.049 & -0.052; -0.042 & -0.040; -0.036 \\
    DDAP & 0.224; 0.240 & 0.148; 0.233 & 0.194; 0.223 \\
    EQeq & -0.119; 0.093 & 0.098; 0.161 & -0.047; -0.033 \\
    \hline
  \end{tabular}
\end{table}

\begin{table}[!h]
   \caption{Range of variation of the atomic partial charges for atom Zn, calculated for the structures DMOF-1-ini, DMOF-1-opti1 and DMOF-1-opti2, using 5 different methods, namely REPEAT, Bader, Mulliken, DDAP, and EQeq.}
   \centering
  \label{tbl:tb2_charges}
  \begin{tabular}{cccc}
    \hline
    Method &  DMOF-1-ini & DMOF-1-opti1 & DMOF-1-opti2\\
    \hline  
    REPEAT & 0.962; 0.968 & 0.881; 0.926 & 0.920; 0.922 \\
    Bader & 1.251; 1.269 & 1.258; 1.285 & 1.074; 1.082 \\
    Mulliken & 0.516; 0.519 & 0.502; 0.505 & 0.565; 0.568 \\
    DDAP & 0.855; 0.856 & 0.810; 0.831 & 0.806; 0.809 \\
    EQeq & 1.092; 1.143 & 1.072; 1.144 & 1.131; 1.132 \\
    \hline
  \end{tabular}
\end{table}

\begin{table}[!h]
   \caption{Range of variation of the atomic partial charges for atom H2, calculated for the structures DMOF-1-ini, DMOF-1-opti1 and DMOF-1-opti2, using 5 different methods, namely REPEAT, Bader, Mulliken, DDAP, and EQeq.}
   \centering
  \label{tbl:tb3_charges}
  \begin{tabular}{cccc}
    \hline
    Method &  DMOF-1-ini & DMOF-1-opti1 & DMOF-1-opti2\\
    \hline  
    REPEAT & 0.048; 0.126 & 0.024; 0.202 & 0.047;	0.147 \\
    Bader & -0.158; 0.112 & -0.258; 0.052 &-0.236;	0.081 \\
    Mulliken &0.067; 0.073 & 0.064; 0.080 & 0.063; 0.070 \\
    DDAP & -0.086; -0.036 & -0.090; 0.004 & -0.087; -0.029 \\
    EQeq & -0.030; 0.083 & -0.039; 0.112 & 0.035; 0.038 \\
    \hline
  \end{tabular}
\end{table}

We have discussed the influence of the method for calculating charges, but even more
interesting is the influence that the geometry of the framework has on the charges. When the
same method is employed, the slight variations of the framework geometry that exist between
the three structures induce significant differences in atomic charges. For example, in Table \ref{tbl:tb1_charges} we
see that the charge of atom C3 calculated with the EQeq method can vary from $-0.119$ to $0.093$
for the DMOF-1-ini structure, but for the DMOF-1-opti1 there are no negatively charged C3
atoms. This is a weak point of the force field-based calculations, which rely upon the validity of
the charges to provide an adequate description of the electrostatic interactions. The influence of
the geometry on the calculated charges is more marked for the Bader and REPEAT methods,
while the Mulliken method seems to be the one that minimises the spread of charges for atoms
that are symmetrically equivalent. The DDAP method also shows and acceptable spread of
charges, and if we take into account both the advantages and drawbacks of the two methods,
discussed in the previous section, we would suggest using these charges for the calculation of
atomic partial charges. If a screening of a large number of structures will be performed the use
of DDAP charges is unfeasible. In that case, the EQeq method provides reasonably good
charges, at a low computational cost, so that method would be the method of choice.

\section{Influence of the chosen charges on structural properties}
The effect of the charges on the calculation of adsorption heats, diffusion constants and
separation properties has already been treated in the literature, as shown in section 1. Here we
discuss how charges affect the structural behavior of MOFs. To do this, we have selected
IRMOF-1, which is known to show a negative thermal expansion \cite{zhou2008origin,dubbeldam2007exceptional}. The atomic
charges reported by \citet{dubbeldam2007exceptional} were scaled by 0.95 and 1.05, and the thermal
behavior was studied by means of molecular dynamics. The framework has been modelled by
molecular dynamics simulations in the isothermal-isobaric (NPT) ensemble (fully flexible cell,
using Nosé-Hoover thermostat and Parrinello-Rahman barostat). Intramolecular interactions
were taken into account employing the force field developed by \citet{dubbeldam2007exceptional}. The
external pressure is set to zero. The simulations have been run for 5 ns, using an integration step
of 0.5 fs. Ewald summation was used to calculate the electrostatic energy in the crystalline
framework, and a cut-off radius of 12 \AA~ was used for short-range interactions. We have used the
RASPA code to carry out the simulations \cite{RASPA_code}.
\begin{figure}[!h]
	\begin{center}
		\centering
	    \includegraphics[width=0.40\textwidth]{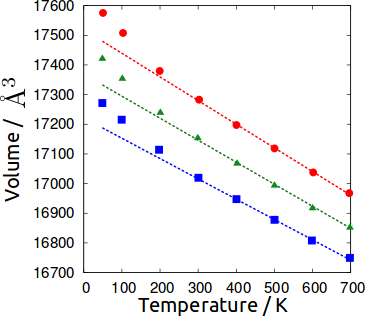}
	    \includegraphics[width=0.38\textwidth]{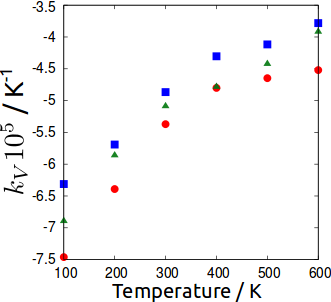}
	    \caption{Left: Dependence of the cell volume with the temperature, for IRMOF-1. Right: Dependence of the thermal expansion coefficient with the temperature, for IRMOF-1.}
	    \label{fig:MOF5_NTE}
	\end{center}
\end{figure}

In Figure \ref{fig:MOF5_NTE}-Left, we show the dependence with temperature of the cell volume, in IRMOF-1, for
three different sets of charges. Since the charges are homogeneously changed in the whole unit
cell, and the charges do not affect the bond strengths, it is somewhat surprising that the small
changes introduced in the charges (5\%) produce a significant modification in the (negative)
thermal expansion of IRMOF-1. For each temperature, it is observed that there is an inverse
dependence of the cell volume with the amount of charge scaling, which is an evidence of the
role of long range (Coulombic) interactions in the overall structure of MOFs. However, the rate
of the structural changes with temperature has a direct dependence with the charges, as revealed
by the behavior of the thermal expansion coefficient (Figure \ref{fig:MOF5_NTE}-Right).
This is probably due to a balance between the elastic and the entropic effects, as long range forces compete with the
bonding interactions that are not modified by the charges.

\section{Conclusions}
We have reviewed the different methods available to calculate atomic partial charges in MOFs,
and we have also presented two examples of materials in which the choice of charges has a big
influence on the results obtained. The decision about what method is the best is not a simple
one, and the choice will depend on factors such as the knowledge and experience of the
researcher, the codes that he or she has access to, the type of systems that will be studied, etc.
Once a method has been chosen, it is important to check carefully that all charges are
chemically sound. And, if possible, it is desirable to compare the charges obtained with more
than one method. We also suggest charge calculations on structures optimized by different
approaches, as small structural differences might have a large impact on the resulting atomic
charges. We have also shown that not only molecular adsorption, separation and diffusion are
affected by the choice of the charges, but also the structural properties, which is particularly
relevant for modeling systems with at least certain degree of flexibility.

\section{Acknowledgments}
The research leading to these results has received funding from the European Research Council under the European Union's Seventh Framework Programme (FP7/2007-2013) / ERC grant agreement nº [279520]), the Spanish \emph{Ministerio de Economía y Competitividad} (CTQ2013-48396-P), and the Netherlands Council for Chemical Sciences (NWO/CW) through a VIDI grant. SRG thanks the \emph{Ministerio de Economía y Competitividad} for his predoctoral fellowship.
\bibliography{biblio}
\end{document}